\documentstyle[psfig,galley, rotate]{mn}

\title[Large-scale Cosmic Homogeneity in the PSCz
Catalogue]{Large-scale Cosmic Homogeneity from a Multifractal
Analysis of the PSCz Catalogue}

\author[Jun Pan\& Peter Coles]{Jun Pan and Peter Coles\\
School of Physics \& Astronomy, University of Nottingham,
University Park, Nottingham, NG7 2RD, United Kingdom\\ }

\begin{document}

\maketitle

\begin{abstract}
We investigate the behaviour of galaxy clustering on large scales
using the PSCz catalogue. In particular, we ask whether there is any
evidence of large-scale fractal behaviour in this catalogue. We find
the correlation dimension in this survey varies with scale, consistent
with other analyses. For example, our results on small and
intermediate scales are consistent those obtained from the QDOT
sample, but the larger PSCz sample allows us to extend the analysis
out to much larger scales. We find firm evidence that the sample
becomes homogeneous at large scales; the correlation dimension of the
sample is $D_2=2.992\pm 0.003$ for $r>30 h^{-1}$ Mpc. This provides
strong evidence in favour of a Universe which obeys the Cosmological
Principle. \end{abstract}


\section{Introduction}
The most fundamental assumption underlying the standard
cosmological models is that, on sufficiently large scales, the
Universe is homogeneous and isotropic. We know that, on small
scales, matter is distributed in a roughly hierarchical fashion
with galaxies being grouped into clusters which in turn are
grouped into superclusters. This has led some to argue for a
radically different cosmological paradigm in which this hierarchy
carries on {\em ad infinitum} (Coleman, Pietronero \& Sanders 1988;
Coleman \& Pietronero 1992). In such a case the Universe is not smooth
on large scales, but has a fractal structure in which the Cosmological
Principle, at least in its usual form, does not apply (Sylos-Labini et
al. 1998). On the other hand, many others have argued that redshift
surveys show a definite transition to large-scale homogeneity (e.g.
Guzzo 1997; Cappi et al. 1998; Mart\'{\i}nez et al. 1998; Scaramella et
al. 1998).

There is still some degree of controversy about the large-scale
homogeneity of the Universe, primarily because the main sources of
direct information on spatial structure, redshift surveys, have either
been too shallow or too sparse or of too restricted a geometry to
provide conclusive evidence. This has led to arguments about the
treatment of boundary effects and other sampling errors, leading to
different groups reporting different results for the same surveys
(e.g. Mart\'{\i}nez \& Jones 1990; Sylos-Labini et al. 1998). Given
these bones of contention, the best arguments in favour of large-scale
homogeneity still stem from the near-isotropy of sources or background
radiation observed in projection on the sky (Wu, Lahav \& Rees 1999).

In this paper we address this issue by performing a multifractal
analysis of the PSCz catalogue of redshifts of IRAS galaxies. This is
a good sample to use for this issue because it is both extremely deep,
allowing truly large-scale structure to be probed) but is also all-sky
(reducing the influence of boundary corrections). The method we use
involves a generalization of the fractal dimension to a set of
generalised dimensions based on the scaling properties of different
moments of the galaxy counts (Mart\'{\i}nez et al. 1995). The work we
describe here represents the culmination of work begun by Mart\'{\i}nez \&
Coles (1994) which analyzed the QDOT catalogue, a one-in-six subsample
of the parent IRAS catalogue (Lawrence et al. 1999).

\section{The Sample}

\begin{figure*}
\psfig{file=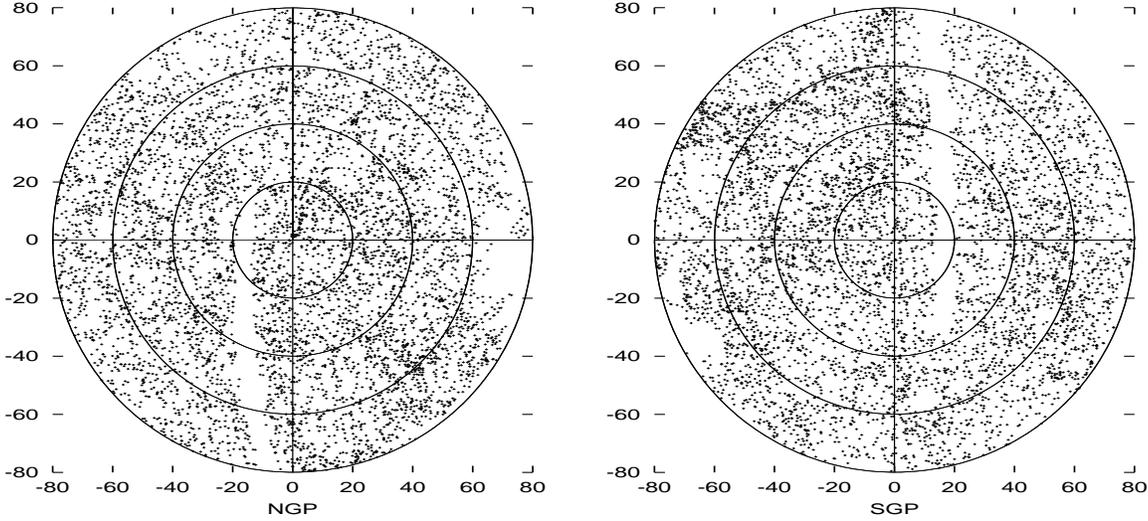,height=7cm,width=8cm,angle=-90}
\caption{The projection of galaxies selected for the sample in the
northern and southern hemispheres, respectively.}
\end{figure*}

The PSCz catalogue consists of $15411$ IRAS galaxies across $84\%$ of
the sky with flux at $60{\mu}m$ greater than $0.60$ Jy. In all, the
redshifts of $14677$ galaxies are available (Saunders et al 2000). We
use a subset of this catalogue, defined as follows. First, we exclude
galaxies with galactic latitudes $|b|<10^{\circ}$ for this analysis to
avoid problems near the galactic plane. In order to minimize the
influence of local structure and large-scale sampling problem, the
distance is limited in the range $10h^{-1}$ Mpc $<R< 400h^{-1}$ Mpc,
where $R$ is calculated from the redshift $z$ via the Mattig formula:
$$ R=\frac{c}{H_0 {q_0}^2 (1+z)}[q_0 z + (q_0 -1)(\sqrt{2 q_0 z
+1}-1)], \eqno(1)
$$
with $H_0=100h^{-1}$ km s$^{-1}$ Mpc$^{-1}$ and
$q_0=0.5$.

As usual, we define the selection function $\phi(R)$ as the
probability that a galaxy at a distance R is included in the
catalogue. An cutoff in absolute luminosity is set in order that
$\phi(R)=1$ when $R<40h^{-1}$ Mpc. For a 60-micron flux limit of
 $S_{60}>0.60$ the luminosity cutoff is $10^{9.19}L_{\odot}$ at
 $40h^{-1}$.
The form of $\phi(R)$ can be derived directly from the parametric
selection function given by Saunders et al (2000). The final sample
has $11901$ galaxies. For reference the distribution of galaxies is
shown in Figure 1, while Figure 2 is the plot of $\phi(R)$ against
$R$.

\begin{figure}
\begingroup%
  \makeatletter%
  \newcommand{\GNUPLOTspecial}{%
    \@sanitize\catcode`\%=14\relax\special}%
  \setlength{\unitlength}{0.1bp}%
\begin{picture}(3600,2160)(0,0)%
\special{psfile=sf llx=0 lly=0 urx=720 ury=504 rwi=7200}
\put(1430,50){\makebox(0,0){$logR(h^{-1}Mpc)$}}%
\put(100,1180){%
\special{ps: gsave currentpoint currentpoint translate
270 rotate neg exch neg exch translate}%
\makebox(0,0)[b]{\shortstack{$\phi(R)$}}%
\special{ps: currentpoint grestore moveto}%
}%
\put(2310,200){\makebox(0,0){3}}%
\put(2017,200){\makebox(0,0){2.5}}%
\put(1723,200){\makebox(0,0){2}}%
\put(1430,200){\makebox(0,0){1.5}}%
\put(1137,200){\makebox(0,0){1}}%
\put(843,200){\makebox(0,0){0.5}}%
\put(550,200){\makebox(0,0){0}}%
\put(500,2060){\makebox(0,0)[r]{10}}%
\put(500,1708){\makebox(0,0)[r]{1}}%
\put(500,1356){\makebox(0,0)[r]{0.1}}%
\put(500,1004){\makebox(0,0)[r]{0.01}}%
\put(500,652){\makebox(0,0)[r]{0.001}}%
\put(500,300){\makebox(0,0)[r]{0.0001}}%
\end{picture}%
\endgroup
 
\caption{Selection Function. The solid line is the expected function
while points with error bars are the actual counting result of the
sample.}
\end{figure}
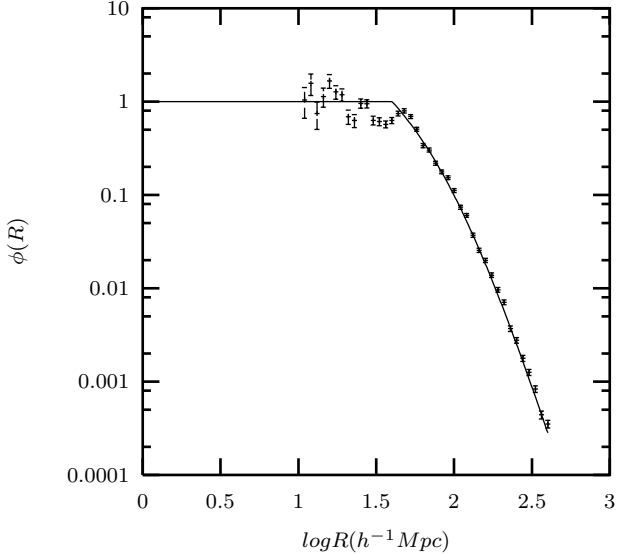

\section{Multifractal Analysis}

The measure we use is constructed from the partition function,
$$
Z(q,r)=\frac{1}{N}\sum_{i=1}^{N} p_i(r)^{q-1}\propto r^{\tau (q)}, \eqno(2)
$$
with $p(i)=n_i(r)/N$,
where $n_i(r)$ is the count of objects in the cell of
radius $r$ centered upon an object labeled by $i$ (which is not
included in the  count). For each value of $q$ in equation (2), one
can have a different scaling exponent of the set $\tau(q)$.

It is necessary in practical  applications such as this to account for
edge effects and selection. We take the corrected local count around
the $i$th object to be
$$ n_i(r)=\frac{1}{f_i(r)} \sum_{j=1}^{N}
\frac{\Psi(|\textbf{r}_j - \textbf{r}_i| - r)}{\phi(r_j)}, \eqno(3)
$$ and
$$
\Psi(x)=\left\{\begin{array}{cc} 1, & x \leq 0\\ 0, & x> 0
\end{array}\right.,\eqno(4)
$$
where $f_i(r)$ is the volume fraction
of the sphere centered on the object of radius $r$ within the boundary
of the sample.

The scaling exponents $\tau(q)$ lead to the definition of the
so-called Renyi dimensions: $$ D_q=\frac{\tau(q)}{q-1}, \eqno(5)
$$ where $D_q$ is the spectrum of fractal dimensions for a fractal
measure on the sample. The $D_q$ for each value of $q$ gives
information about the scaling properties of different aspects of
the density field. For high $q$, $D_q$ tells us about high-density
regions while for low $q$ (including negative values) the measure
is weighted towards low-density regions. For $q=1$, $D_1$ can be
derived from $$ S(r)= \frac{1}{N}\sum_{i=1}^{N}\log p_i(r) \propto
r^{D_1}, \eqno(6) $$ where $S(r)$ is the partition {\em entropy}
of the measure on the sample set; $D_1$ is consequently termed the
information dimension.

Of more direct interest in this case is the case $q=2$. The
exponent $D_2$ is what is generally called the correlation
dimension, and it is related to the usual two-point correlation
function $\xi(r)$ for a sample displaying large-scale homogeneity
(Peebles 1980). If the mean number of neighbours around a given point
is $\langle n \rangle$ then
$$
\langle n \rangle=4\pi\bar{n} \int_{0}^{r} [1+\xi(r)]s^2 ds.
\eqno(7)
$$
In this case $\langle n \rangle  \sim r^{\alpha}$ means $\alpha=D_2$. A
homogeneous distribution has $D_2=3$, whereas a power-law in
$1+\xi(r)\sim r^{-\gamma}$ yields $D_2=3-\gamma$.

\section{Results and Discussion}
We illustrate the results in Figure 3 by plotting $D_q$ against
cell size $R$ for $q=2$. Considering the case $q=2$ first, it is
interesting to compare the results with those obtained for the
QDOT sample (Mart\'{\i}nez \& Coles 1994). For $r$ below $\sim
10h^{-1}$ Mpc, we get $D_2=2.16$. The QDOT value was 2.25. Above
$r \sim 30h^{-1}$ Mpc, $D_2=2.99$ which closely approaches the
value $D=3$ for homogeneous distribution. The formal error for the fit
is around $0.003$, of similar size to the Poisson error.

Within the range  $10h^{-1}$ Mpc to $30h^{-1}$ Mpc, there is an
intermediate regime represented by a gradual transitions from
fractal to homogeneous behaviour. To compare with the range quoted by
Mart\'{\i}nez \& Coles (1994), we fit this intermediate regime
($10h^{-1}$ Mpc $<r <$ $50h^{-1}$ Mpc) and obtained $D_2=2.71$,
consistent with the value $D_2= 2.77$ obtained for the QDOT catalogue.
Notice that, on small scales, the value of $D_2$ will be affected by
peculiar motions since we work entirely in redshift space, but this is
not expected to be the case on large scales.
\begin{figure}
\psfig{file=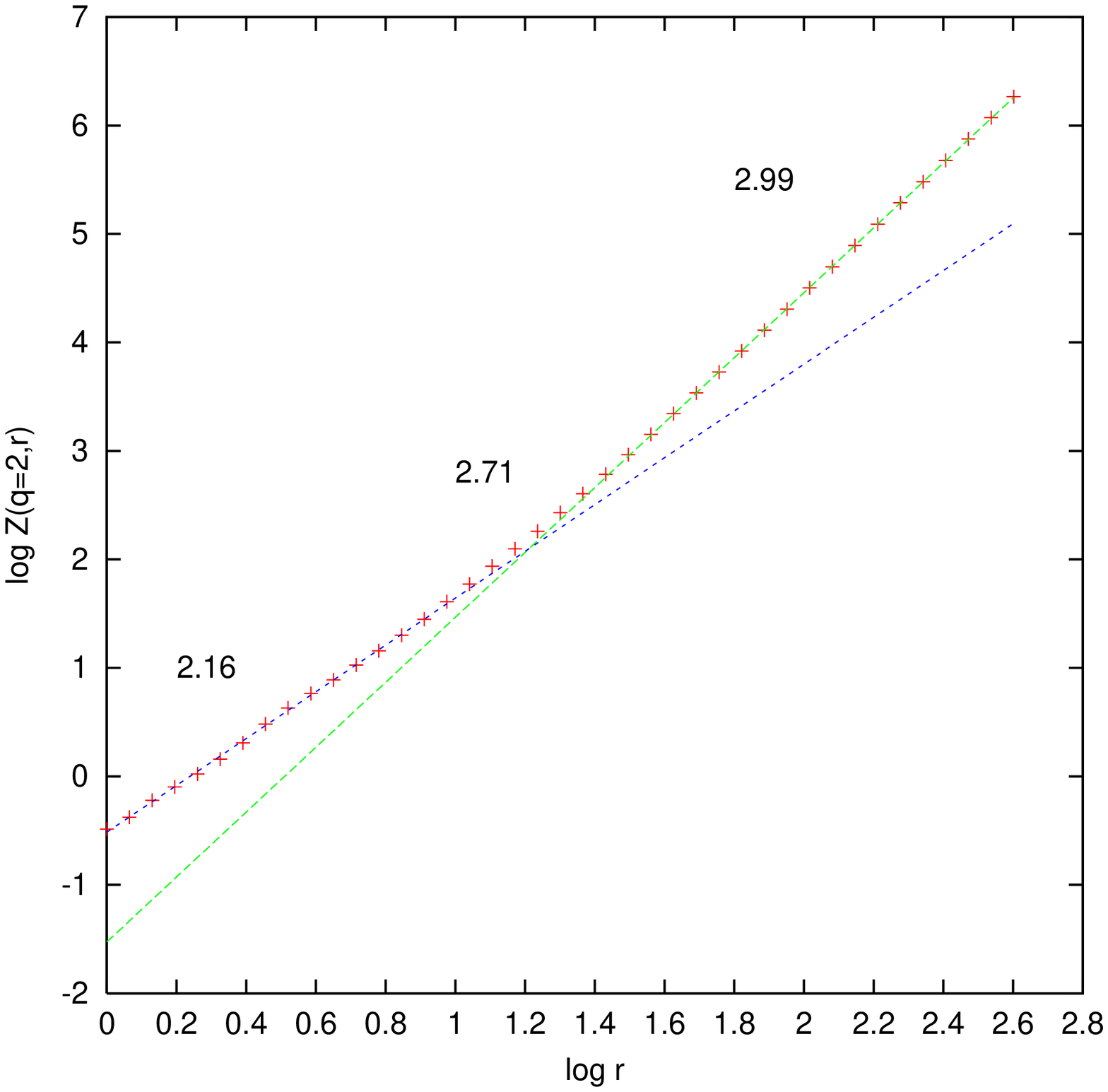,height=110mm,width=110mm,angle=-90}
\psfig{file=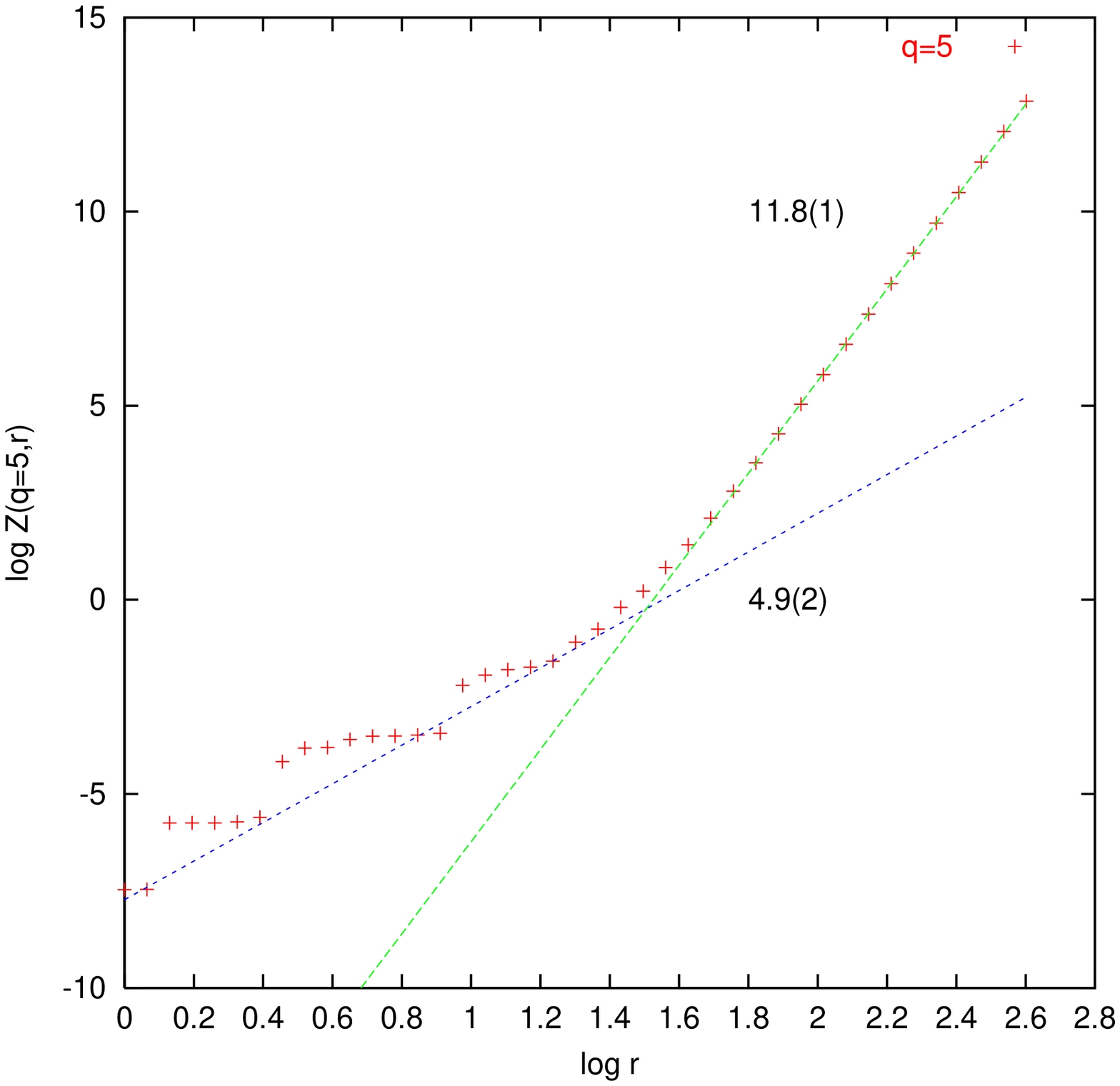,height=110mm,width=110mm,angle=-90}
\caption{$\log Z(q,r)\:vs.\:\log r$ for $q=2$ and $q=5$. The fit for
intermediate scales is not plotted.}
\end{figure}

We show the results for $q=5$ in order to illustrate some of the
difficulties with estimating $D_q$ for large $q$. Notice that for
scales larger than $\sim 50h^{-1}$ Mpc, $\tau(q)=11.8$, so $D_5=2.95$.
This is consistent with the tendency to homogeneity discussed in the
previous paragraph. However, on smaller scales (say below $r \sim
20h^{-1}$ Mpc), the partition function displays a series of steps as a
function of  $r$. This systematic tendency is clearer in the PSCz
sample than in QDOT and is due to the suppression of high $q$ to the
low value of $n_i(r)$ at certain levels. The end points of the steps
probably correspond to some characteristic scales of the hierarchical
structures at different scales. A  possible {\em primary} scaling law
could be recovered by only adopting the end points of each step. This
fitting shows that $\tilde{D_5}=1.23$. Between $20h^{-1}$ and
$50h^{-1}$ Mpc, $D_5=2.05$ in comparison with the value of $2.30$
obtained for QDOT.
\begin{figure}
\psfig{file=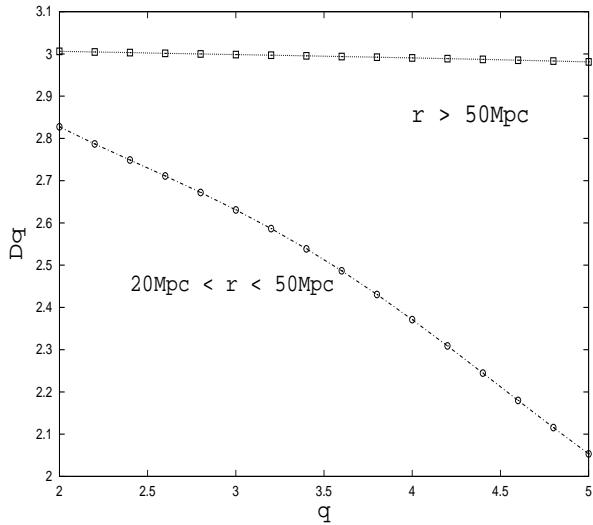,height=70mm,width=80mm,angle=-90}
\caption{The generalized dimensions $D_q$ ($q>2$) for
$r>50h^{-1}$ Mpc and $r$ between $20$ and $50 h^{-1}$ Mpc respectively.}
\end{figure}

The spectrum $D_q$ for $q>2$ is shown in Figure 4. The measure of
Equation (2) is not capable of generating a complete spectrum
including $q<2$ for a point set because of discreteness effects. In
order to cover the latter range, an alternative measure is used:
$$
W[\tau(q),n]=\frac{1}{N} \sum_{i=1}^{N}r_i(n)^{-\tau}\propto
n^{1-q}, \eqno(8)
$$
where $r_i(n)$ is the radius of the smallest
sphere centered at an object that encloses $n$ neighbors. We shall
return to the case $q<2$ in future work.

\section{Conclusions}
The results we have presented here provide very strong evidence
that the distribution of IRAS galaxies becomes homogeneous on
large scales. This is reassuring for adherents of the standard
cosmological framework but runs counter to the alternative,
fractal paradigm.

Our future work on this problem will follow two principal
directions. One is to investigate thoroughly the role of boundary
corrections and sampling in the estimation of $D_q$, including $q<2$,
using simulated catalogues. Our work along these lines so far has
shown that the large-scale value of $D_2$ is robust to changes in
boundary correction, at least for this catalogue, so we can be
confident about its extremely small error bar.

The other important thread is to connected the scaling behaviour
of the matter distribution, in terms of $D_q$, to the
gravitational dynamics of structure formation. Simple scaling
arguments have already born considerable fruit in the
interpretation of large-scale clustering (e.g. Hamilton et al.
1991). With the imminent arrival of even larger surveys, the use
of more sophisticated descriptors will allow a deeper
understanding of how spatial pattern arises in the galaxy
distribution.

\section*{Acknowledgment}
Jun Pan receives an ORS award from the CVCP. We thank Ed Hawkins
for helpful discussions.

\end{document}